\documentclass[english,pra,twocolumn]{revtex4}
\usepackage[T1]{fontenc}
\usepackage[latin9]{inputenc}
\usepackage{verbatim}
\usepackage{booktabs}
\usepackage{amsmath}
\usepackage{graphicx}

\providecommand{\tabularnewline}{\\}

\usepackage{babel}

\begin{document}

\title{Quantum Algorithm for Molecular Properties and Geometry Optimization}

\author{Ivan Kassal}

\email{kassal@fas.harvard.edu}

\affiliation{Department of Chemistry and Chemical Biology, Harvard University,
Cambridge, MA 02138}

\author{Al{\'a}n Aspuru-Guzik}

\email{aspuru@chemistry.harvard.edu}

\affiliation{Department of Chemistry and Chemical Biology, Harvard University,
Cambridge, MA 02138}

\date{\today}

\begin{abstract}
It is known that quantum computers, if available, would allow an exponential
decrease in the computational cost of quantum simulations. We extend
this result to show that the computation of molecular properties (energy
derivatives) could also be sped up using quantum computers. We provide
a quantum algorithm for the numerical evaluation of molecular properties,
whose time cost is a constant multiple of the time needed to compute
the molecular energy, regardless of the size of the system. Molecular
properties computed with the proposed approach could also be used
for the optimization of molecular geometries or other properties.
For that purpose, we discuss the benefits of quantum techniques for
Newton's method and Householder methods. Finally, global minima for
the proposed optimizations can be found using the quantum basin hopper
algorithm, which offers an additional quadratic reduction in cost
over classical multi-start techniques.
\end{abstract}

\maketitle

Applying \emph{ab initio} methods of quantum chemistry to particular
problems often requires computing derivatives of the molecular energy.
For instance, obtaining a molecule's electric properties relies on
the ability to compute derivatives with respect to external electromagnetic
fields. Likewise, computing the gradient of the molecular energy with
respect to the nuclear coordinates is the most commonly used method
for the proper characterization of potential energy surfaces and for
optimizing the geometry of all but the smallest molecules. The computation
of these kinds of derivatives, known as \emph{molecular properties},
is nowadays a routine matter when it comes to low-order derivatives
or small systems (or both). This is largely due to advances in analytical
gradient techniques, which allow for explicit property evaluation
without resorting to numerical differentiation \citep{pulay_ab_1969,pulay_analytical_1987,helgaker_analytical_1988,yamaguchi_new_1994,shepard_analytic_1995,pulay_analytical_1995,helgaker_1998}. 

Nevertheless, the computation of higher-order derivatives is often
prohibitively expensive, even though such derivatives are often needed.
For example, third- and fourth-order anharmonic constants are sometimes
required to accurately compute a vibrational absorption spectrum \citep{helgaker_analytical_1988}
or efficiently determine the location of transition states on complex
potential energy surfaces \citep{pulay_analytical_1995}. Other properties
of interest, such as hyperpolarizabilities, Raman intensities, or
vibrational circular dichroism, are also cubic or quartic derivatives.
In this report, we show that quantum computers, once available, will
be able to bypass some of the high cost of computing these properties.
In particular, we show that any molecular property can be evaluated
on a quantum computer using resources that, up to a small constant,
are equal to those required to compute the molecular energy once. 
We have previously characterized the
advantage of quantum computers at both computing molecular energies
\citep{aspuru-guzik_simulated_2005,wang_quantum_2008} and simulating
chemical reaction dynamics \citep{kassal_polynomial-time_2008}, and
the present work extends our program to molecular properties.

This paper begins with a brief overview of classical techniques for
the evaluation of molecular properties, both numerical and analytical.
We then introduce the quantum algorithm for molecular properties,
and discuss its advantages and disadvantages with respect to classical
techniques. We conclude with geometry optimization as a particular
example, and we show that it can benefit from an additional quadratic
speed-up through Grover's search \citep{grover_fast_}.

\section{The classical methods}

Given an external perturbation $\boldsymbol{\mu}$, the total molecular
electronic energy can be expanded in a Taylor series\begin{equation}
E(\boldsymbol{\mu})=E^{(0)}+\boldsymbol{\mu}^{\top}\mathbf{E}^{(1)}+\frac{1}{2}\boldsymbol{\mu}^{\top}\mathbf{E}^{(2)}\boldsymbol{\mu}+\ldots\end{equation}
where the coefficients $\mathbf{E}^{(n)}$ are called the \emph{molecular
properties} and describe the response of the system to the applied
perturbation \citep{helgaker_1998}. We consider time-independent
properties, which can be obtained by differentiating the energy at
$\boldsymbol{\mu}=\mathbf{0}$,\begin{equation}
\mathbf{E}^{(n)}=\left.\frac{\text{d}^{n}E}{\text{d}\boldsymbol{\mu}^{n}}\right|_{\mathbf{0}}.\end{equation}
Many examples of useful derivatives can be given. For instance, the
derivatives with respect to the electric field $\mathbf{F}$ are the
permanent electric dipole, the static polarizability, and the static
hyperpolarizabilities of various orders:\begin{equation}
\left.\frac{\text{d}E}{\text{d}\mathbf{F}}\right|_{\mathbf{0}}=-\mathbf{d},\;\left.\frac{\text{d}^{2}E}{\text{d}\mathbf{F}^{2}}\right|_{\mathbf{0}}=-\boldsymbol{\alpha},\;\left.\frac{\text{d}^{3}E}{\text{d}\mathbf{F}^{3}}\right|_{\mathbf{0}}=-\boldsymbol{\beta},\ldots\end{equation}
where the subscript denotes differentiation at $\mathbf{F}=\mathbf{0}$.
The derivatives with respect to nuclear coordinates $\mathbf{R}$
include the forces on the nuclei and the force constants, while mixed
derivatives can provide information such as Raman intensities \citep{pulay_analytical_1987}.

On a classical computer, an energy derivative can be evaluated either
numerically or analytically, and we discuss each approach in turn.

\begin{figure}
\includegraphics[scale=0.55]{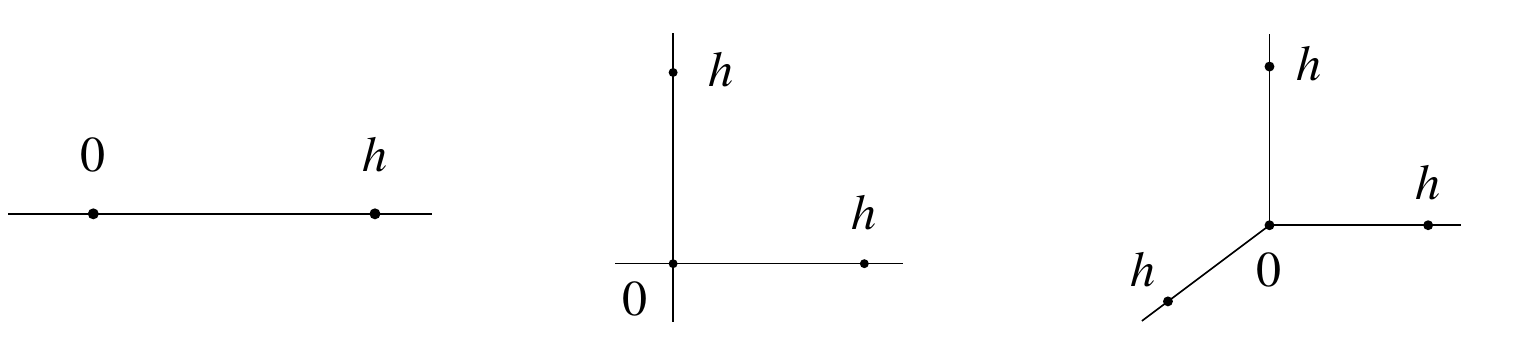}

\caption{Obtaining a numerical gradient of a function defined on a $d$-dimensional
space classically requires sampling the function $d+1$ times, once
at the origin and once at a distance $h$ along each of the axes.
Shown above are the sample points for the cases $d=1$ through $d=3$.
The quantum gradient algorithm can evaluate many sample points in
superposition, producing the same calculated gradient using only
one call to the function.\label{fig:dimensions}}

\end{figure}

Numerical derivative techniques rely on computing the value of the
energy at several discrete points, and then using those values to
estimate the true derivative. The simplest technique is finite difference,
which for the first derivative in one dimension is the familiar formula,\begin{equation}
\left.\frac{\text{d}E}{\text{d}\mu}\right|_{0}\approx\frac{E(h)-E(0)}{h}.\end{equation}
In $d$ dimensions, computing the gradient requires at least $d+1$
evaluations of the energy, once at the origin and once at a distance
$h$ along each axis (Fig. \ref{fig:dimensions}). Similarly, evaluating
higher-order derivatives requires the knowledge of the energy on a
particular grid, with at least $d^{n}+1$ points for the $n^{\mathrm{th}}$
derivative.

While numerical gradient techniques usually require minimal effort
to implement, they are occasionally susceptible to numerical instability,
due to the ill-posedness of numerical differentiation in general \citep{tikhonov_solutions_1977}.
This is particularly problematic when using finite-precision arithmetic,
where various rounding errors can accumulate and be amplified upon
division by the small number $h$. The fact that small errors in the
evaluated function can lead to large errors in the derivative affects
\emph{ab initio} electronic structure methods insofar as they usually
involve long calculations with many potential sources of error, including
rounding and quadrature.

By contrast, analytic derivative techniques are those that compute
the derivative by direct evaluation of an analytic expression. They
were introduced in quantum chemistry by Pulay \citep{pulay_ab_1969},
and have since largely supplanted numerical procedures. They are numerically
more stable and, more importantly, they are usually faster as well.

Analytic gradient formulas exist for just about all electronic structure
techniques and for most kinds of perturbations. To illustrate the
argument and establish the correct scaling, we will describe the particularly
simple case of derivatives of fully variational wavefunctions. We
start by writing the molecular energy as a function $E(\boldsymbol{\mu},\boldsymbol{\lambda}(\boldsymbol{\mu}))$
of the external perturbation $\boldsymbol{\mu}$ and the wavefunction
parameters $\boldsymbol{\lambda}(\boldsymbol{\mu})$. These parameters,
such as the configuration interaction coefficients, completely describe
the electronic wavefunction. Although $\boldsymbol{\lambda}(\boldsymbol{\mu})$
is a function of $\boldsymbol{\mu}$, for simplicity we will write
only $\boldsymbol{\lambda}$. The energy is said to be \emph{fully
variational} with respect to $\boldsymbol{\lambda}$ if, for any given
$\boldsymbol{\mu}$, $\boldsymbol{\lambda}$ assumes the value $\boldsymbol{\lambda}^{*}$
such that the variational condition holds:\begin{equation}
\left.\frac{\partial E(\boldsymbol{\mu},\boldsymbol{\lambda})}{\partial\boldsymbol{\lambda}}\right|_{*}=\mathbf{0},\end{equation}
where $*$ indicates $\boldsymbol{\lambda}=\boldsymbol{\lambda}^{*}$.
In that case we can write $E(\boldsymbol{\mu})=E\left(\boldsymbol{\mu},\boldsymbol{\lambda}^{*}\right)$. 

For fully variational wavefunctions, the gradient with respect to
$\boldsymbol{\mu}$ is given by\begin{multline}
\frac{\text{\text{d}}E(\boldsymbol{\mu})}{\text{d}\boldsymbol{\mu}}=\left.\frac{\partial E(\boldsymbol{\mu},\boldsymbol{\lambda})}{\partial\boldsymbol{\mu}}\right|_{*}+\left.\frac{\partial E(\boldsymbol{\mu},\boldsymbol{\lambda})}{\partial\boldsymbol{\lambda}}\right|_{*}\frac{\partial\boldsymbol{\lambda}}{\partial\boldsymbol{\mu}}=\\
=\left.\frac{\partial E(\boldsymbol{\mu},\boldsymbol{\lambda})}{\partial\boldsymbol{\mu}}\right|_{*}=\left\langle \boldsymbol{\lambda}^{*}\left|\frac{\partial H}{\partial\boldsymbol{\mu}}\right|\boldsymbol{\lambda}^{*}\right\rangle \end{multline}
where we have used the variational condition and the Hellman-Feynman
theorem. Since one need not know the first-order wavefunction response
$\frac{\partial\boldsymbol{\lambda}}{\partial\boldsymbol{\mu}}$,
computing the gradient is, to within a small constant factor \citep{pulay_analytical_1995},
as hard as computing the energy. That is, once $\left|\boldsymbol{\lambda}^{*}\right\rangle $
is available, calculating the expectation value of the Hamiltonian
has approximately the same computational cost as calculating its derivative.
However, computing the second derivative matrix does require the knowledge
of the first-order wavefunction response. In fact, as a direct consequence
of Wigner's $2n+1$ rule of perturbation theory, one needs to know
the first $n$ wavefunction responses in order to calculate the $(2n+1)^{\text{th}}$
derivative. Computing the responses often becomes the bottleneck,
and it is what leads to a higher asymptotic cost of higher-order derivatives.
While the gradient requires about the same resources as the energy,
the second and third derivatives require resources that scale as $O(d)$
times the cost of computing the energy (where $d$ is the number of
degrees of freedom, i.e., the dimension of $\boldsymbol{\mu}$) \citep{pulay_analytical_1995}.
This scaling comes about because $O(d)$ time is required to compute
the matrix $\frac{\partial\boldsymbol{\lambda}}{\partial\boldsymbol{\mu}}$.
Likewise, the scaling of the $(2n+1)^{\text{th}}$ derivative is $O(d^{n})$,
because the bottleneck becomes the computation of the $n^{\text{th }}$
order wavefunction response. In other words, the computational cost
of finding the $n^{\mathrm{th}}$ analytical derivative is $O(d^{\left\lfloor n/2\right\rfloor })$,
roughly a quadratic speed-up over the $O(d^{n})$ numerical methods
of the same degree.

The fact that the scaling of derivative techniques, both numerical
and analytical, depends on $d$ has meant that these techniques are
often restricted to small systems %
\footnote{For some classes of useful properties, $d$ is independent of system
size. For example, if the perturbation is the electric field, then
$d=3$, and indeed there are classical techniques for computing electrical
properties of large molecules. The quantum speed-up is therefore most
pronounced in cases where $d$ varies with system size, as it does
whenever there is differentiation with respect to nuclear coordiates.%
}. This is most acutely true of the molecular Hessian, which is often
beyond reach, even though the gradient is routinely accessible. We
now show that if quantum computers were available, the cost of the
higher derivatives would no longer be prohibitive.

\section{The quantum algorithm}

The quantum algorithm for molecular properties is based on Jordan's
quantum gradient estimation algorithm \citep{jordan_fast_2005}.
Jordan's method can numerically compute the gradient of any function
$F$, given a black box (oracle) that computes the value of $F$ for an arbitrary
input. In particular, the algorithm can evaluate the gradient using
a \emph{single} query to $F$, regardless of the number of dimensions
$d$ of the domain of $F$. By contrast, the simplest classical finite-difference
scheme would require $d+1$ queries to $F$ (see Fig. \ref{fig:dimensions}).
The speed-up is essentially achieved by being able to sample along
all the $d$ dimensions in superposition.
We apply Jordan's algorithm to the computation of molecular properties
by specifying a way to compute the energy on a quantum computer as
well as by outlining how to obtain higher derivatives. In this section,
we describe the algorithm, its application to quantum chemistry, and
finally argue that a return to numerical techniques for molecular
properties would be justified if quantum computers became feasible.

\begin{table}
\begin{tabular}{cccc}
\toprule 
 & \multicolumn{2}{c}{Classical} & Quantum\tabularnewline
\midrule
Derivative & Numerical & Analytical & Numerical\tabularnewline
\midrule
$\frac{\text{d}E}{\text{d}\boldsymbol{\mu}}$ & $d+1$ & $O\left(1\right)$ & $1$\tabularnewline
$\frac{\text{d}^{2}E}{\text{d}\boldsymbol{\mu}^{2}}$ & $d^{2}+1$ & $O\left(d\right)$ & $2$\tabularnewline
$\frac{\text{d}^{3}E}{\text{d}\boldsymbol{\mu}^{3}}$ & $d^{3}+1$ & $O\left(d\right)$ & $4$\tabularnewline
$\vdots$ & $\vdots$ & $\vdots$ & $\vdots$\tabularnewline
$\frac{\text{d}^{n}E}{\text{d}\boldsymbol{\mu}^{n}}$ & $d^{n}+1$ & $O\left(d^{\left\lfloor n/2\right\rfloor }\right)$ & $2^{n-1}$\tabularnewline
\bottomrule
\end{tabular}

\caption{Time resources required by various techniques of computing molecular
properties, in terms of the cost of computing the energy. For example,
the entry ``$d+1$'' means that computing the property requires
$d+1$ evaluations of the molecular energy, while the entries in the
``Analytical'' column indicate comparable computational effort.
$E$ is the total electronic energy, $\boldsymbol{\mu}$ is the external
perturbation, and $d$ is the dimension of $\boldsymbol{\mu}$. All
the derivatives are evaluated at $\boldsymbol{\mu}=\mathbf{0}$. On
classical computers, the numerical scalings correspond to the simplest
finite-difference scheme. Analytical techniques are the ones that
evaluate the derivative directly (the exponent $\left\lfloor n/2\right\rfloor $
comes from Wigner's $2n+1$ rule). On a quantum computer, the quantum
gradient estimation algorithm is used. It should be noted that on
a quantum computer, the number of evaluations of $E$ needed for any
derivative is independent of $d$, and thus of the size of the system.}

\end{table}

We assume that the molecular energy is a smooth, bounded function of the perturbation,
$E:\left[-\frac{h}{2},\frac{h}{2}\right]^{d}\rightarrow\left[E_{\mathrm{min}},E_{\mathrm{max}}\right]$,
where a small $h$ is chosen so that $E$ varies sufficiently slowly over the domain.
For convenience, we express the perturbations in units such that $h$ is unitless
and such that the bounds are the same along all of the axes. We also assume
that we have a black box for $E$, which, given a perturbation $\boldsymbol{\mu}$,
outputs the energy $E(\boldsymbol{\mu})$. The precise nature of the algorithm
inside the black box is not important, so long as it can be implemented on
a quantum computer. In particular, any classical technique of electronic-structure
theory can be converted into a quantum algorithm \citep{nielsen_quantum_2000}.
In Sec. \ref{sec:blackbox}, we will discuss possible realizations of the
black box, including the use of quantum simulation algorithms, which offer a
significant improvement over classical ones.

We begin by choosing the number $n$ of bits of precision that we desire in the
final gradient. Jordan's algorithm starts in an equal superposition on $d$ registers of $n$
qubits each ($nd$ qubits total) \citep{nielsen_quantum_2000}:\begin{equation}
\frac{1}{\sqrt{N^{d}}}\sum_{k_{1}=0}^{N-1}\cdots\sum_{k_{d}=0}^{N-1}\left|k_{1}\right\rangle \cdots\left|k_{d}\right\rangle =\frac{1}{\sqrt{N^{d}}}\sum_{\mathbf{k}}\left|\mathbf{k}\right\rangle ,\end{equation}
where $N=2^{n}$, the states $\left|k_{i}\right\rangle$ are integers
on $n$ qubits represented in binary, and $\left|\mathbf{k}\right\rangle$
is a $d$-dimensional vector of all the $\left|k\right\rangle$'s.

We use this state as an input for the black box for $E$, which will,
for every integer-vector $\left|\mathbf{k}\right\rangle$ in the superposition,
append a phase proportional to the energy $E(\boldsymbol{\mu})$ at
perturbation $\boldsymbol{\mu}=h(\mathbf{k}-\mathbf{N}/2)/N$
(where $\mathbf{N}$ is the vector $(N,N,\ldots,N)$ and serves to center the
sampling region on the origin). To achieve maximum
precision with fewest qubits, one needs an estimate $m$ of the largest magnitude
of any of the first derivatives of $E$. Then, the energy evaluated by the
black box is scaled by a factor $2\pi N/hm$. Because the black box operates on all the
terms in the superposition at once, a single call results in the state \begin{multline}
\frac{1}{\sqrt{N^{d}}}\sum_{\mathbf{k}}\exp\left[\frac{2\pi iN}{hm}E\left(\frac{h}{N}\left(\mathbf{k}-\mathbf{N}/2\right)\right)\right]\left|\mathbf{k}\right\rangle \approx\\
\approx\frac{1}{\sqrt{N^{d}}}\sum_{\mathbf{k}}\exp\left[\frac{2\pi iN}{hm}\left(E(\mathbf{0})+\frac{h}{N}\left(\mathbf{k}-\mathbf{N}/2\right)\cdot\left.\frac{\text{d}E}{\text{d}\boldsymbol{\mu}}\right|_{\mathbf{0}}\right)\right]\left|\mathbf{k}\right\rangle .\end{multline}
The neglect of terms quadratic in $h$ and higher is a valid approximation
for sufficiently small $h$ (the error caused by higher-order terms is
discussed in \citep{jordan_fast_2005} and is only polynomial).
The final state is separable and equals \begin{multline}
\frac{e^{i \Phi(\mathbf{0})}}{\sqrt{N^{d}}} \sum_{k_{1}=0}^{N-1}\exp\left[\frac{2\pi i}{m}k_{1}\left.\frac{\partial E}{\partial\mu_{1}}\right|_{\mathbf{0}}\right]\left|k_{1}\right\rangle\cdots \\
\cdots\sum_{k_{d}=0}^{N-1}\exp\left[\frac{2\pi i}{m}k_{d}\left.\frac{\partial E}{\partial\mu_{d}}\right|_{\mathbf{0}}\right]\left|k_{d}\right\rangle ,\end{multline}
with phase \begin{equation}
\Phi(\mathbf{0})=2\pi\left(\frac{N}{hm}E(\mathbf{0})-\frac{\mathbf{N}}{2m}\cdot\left.\frac{\text{d}E}{\text{d}\boldsymbol{\mu}}\right|_{\mathbf{0}}\right).\end{equation}

Applying the inverse quantum Fourier transform \citep{nielsen_quantum_2000}
to each of the $d$ registers results in the gradient \begin{equation}
e^{i \Phi(\mathbf{0})}\left|\frac{N}{m}\left.\frac{\partial E}{\partial\mu_{1}}\right|_{\mathbf{0}}\right\rangle \cdots\left|\frac{N}{m}\left.\frac{\partial E}{\partial\mu_{d}}\right|_{\mathbf{0}}\right\rangle
=e^{i \Phi(\mathbf{0})}\left|\frac{N}{m}\left.\frac{\text{d}E}{\text{d}\boldsymbol{\mu}}\right|_{\mathbf{0}}\right\rangle.\label{eq:gradient} \end{equation}
The scaling factor $N/m$ ensures that $\frac{N}{m}\frac{\text{d}E}{\text{d}\boldsymbol{\mu}}$
is an integer-vector with $n$ bits of precision along each dimension.
It should be reiterated that a single call to $E$ was made, as opposed
to the $d+1$ that would be needed in the case of numerical
differentiation by finite difference.

Overall, the gradient estimation algorithm produces the transformation\begin{equation}
\left|\mathbf{0}\right\rangle \longrightarrow e^{i \Phi(\mathbf{0})}\left|\frac{N}{m}\left.\frac{\text{d}E}{\text{d}\boldsymbol{\mu}}\right|_{\mathbf{0}}\right\rangle .\end{equation}
We can compute the Hessian (and higher derivatives) by iterating this
algorithm. If, instead of making a call to $E(\boldsymbol{\mu})$,
the algorithm sought $E(\boldsymbol{\mu}-\boldsymbol{\nu})$ from the black box,
we would perform, at the cost of this single additional subtraction,
\begin{equation}
\left|\mathbf{0}\right\rangle \longrightarrow e^{i \Phi(\boldsymbol{\nu})}\left|\frac{N}{m}\left.\frac{\text{d}E}{\text{d}\boldsymbol{\mu}}\right|_{\boldsymbol{\nu}}\right\rangle ,\end{equation}
with global phase
\begin{equation}
\Phi(\boldsymbol{\nu})=2\pi\left(\frac{N}{hm}E(\boldsymbol{\nu})-\frac{\mathbf{N}}{2m}\cdot\left.\frac{\text{d}E}{\text{d}\boldsymbol{\mu}}\right|_{\boldsymbol{\nu}}\right).\end{equation}
Because we will be using this procedure as a subroutine, it is important
to remove (or ``uncompute'') the global phase, which would otherwise
become a relative phase. One additional evaluation of $E$ (at $\boldsymbol{\nu}$)
suffices for this uncomputation. Overall, this supplies
another black box, which, given $\boldsymbol{\nu}$, yields $\left|\frac{N}{m}\left.\frac{\text{d}E}{\text{d}\boldsymbol{\mu}}\right|_{\boldsymbol{\nu}}\right\rangle $
using only two calls to the original black box for $E$. One can use
the gradient algorithm with this new black box, producing the state\begin{equation}
e^{i\Phi^{(2)}(\mathbf{0})}\left|\frac{N}{m^{(2)}}\left.\frac{\text{d}^{2}E}{\text{d}\boldsymbol{\mu}^{2}}\right|_{\mathbf{0}}\right\rangle ,\end{equation}
which is a two-dimensional array of $d^{2}$ quantum registers containing
all the elements of the Hessian matrix of $E$. In addition, $m^{(2)}$ is
an estimate for the magnitude of the largest second derivative, and
the phase is \begin{equation}
\Phi^{(2)}(\mathbf{0})=2\pi\left(\frac{N}{hm^{(2)}}\left.\frac{\text{d}E}{\text{d}\boldsymbol{\mu}}\right|_{\mathbf{0}}-\frac{\mathbf{N}}{2m^{(2)}}\cdot\left.\frac{\text{d}^2 E}{\text{d}\boldsymbol{\mu}^2}\right|_{\mathbf{0}}\right).\end{equation}

Computing higher derivatives
would require additional factors of two in the number of required
black box calls, caused by the need to uncompute a global phase at each
step (this problem is a common feature when it comes to recursing
quantum algorithms \citep{aaronson_quantum_2002}). Hence, evaluating
the $n^{\mathrm{th}}$ derivative requires $2^{n-1}$ queries to $E$,
which, although exponential in $n$, is much better than $d^{n}+1$,
which is the minimum number of function queries required to compute
the derivative by classical finite difference. We stress that the
number of calls to $E$ is independent of $d$, and thus of the size
of the system, for the derivative of any order.

One could remark that the quantum gradient algorithm is a numerical
approach and that therefore, just like classical numerical techniques,
it would be affected by numerical instability. This implies that the
quantum gradient algorithm cannot be used indiscriminately for problems
that feature errors that cannot be controllably reduced through additional
computational effort. For example, finite-precision arithmetic presents
the same problems to quantum computers as it does to classical ones,
but the rounding errors can be brought under control by using more
digits of precision (as on classical computers). Quantum chemistry
techniques might present numerical problems as well, insofar as they
contain uncontrolled sources of error. However, if the technique for
computing the energy is numerically exact, that is to say, if the
error in the energy can be controllably reduced below any level, the
magnitude of the numerical error in the calculated derivative can
likewise be made arbitrarily small. Fortunately, quantum computers
would make it possible to efficiently evaluate the numerically exact
molecular energy, meaning that numerical instability will not be a
problem. We turn to the topic of molecular energy evaluation next.

\section{The black box for the energy}
\label{sec:blackbox} 

The application of Jordan's gradient algorithm to chemical problems
requires that there be a black box that can compute the value of the
ground-state molecular energy at any value of the perturbation $\boldsymbol{\mu}$
in the neighborhood of $\boldsymbol{\mu}=\mathbf{0}$. Furthermore,
to avoid numerical artifacts, this black box should be numerically
exact, allowing the error in the energy to be controllably reduced
through additional computational work.

The problem of exact classical electronic structure methods is that
they generally have a computational cost that scales exponentially
with the size of the system. Although these classical algorithms could
also be used as subroutines in the quantum gradient algorithm, there
are quantum electronic-structure algorithms that could avoid the
exponential scaling in many cases.

In particular, we have recently described a quantum full CI algorithm
\citep{aspuru-guzik_simulated_2005} for computing the molecular ground
state energy to a given precision in $O\left(M^{5}\right)$ time \citep{lanyon_towards_2009},
where $M$ is the number of basis functions. This algorithm could
be easily recast as a subroutine that would function as the black
box for the energy. Several modifications would have to be made, including
the direct computation of
the overlap integrals on the quantum computer, rendering it possible
to introduce the perturbation $\boldsymbol{\mu}$ into the calculation.
Nevertheless, a quantum computer running the quantum FCI algorithm
could be used to obtain a molecular property of a system with basis
size $M$ in $O\left(M^{5}\right)$ time, a dramatic improvement over
the possibilities of classical computers.

A more recent development is the real-space chemical dynamics simulation
algorithm \citep{kassal_polynomial-time_2008,lidar_calculatingthermal_1999},
based on Zalka's earlier work \citep{christof_zalka_simulating_1998}.
It is known that simulating, to a given precision, the exact dynamics
of a system of $P$ particles interacting under a pairwise interaction
requires at most $O(P^2)$ time and $O(P)$ space, in contrast to
the classical exponential cost. If an eigenstate of the system Hamiltonian
were prepared as the initial state \citep{ward_preparation_2009},
the dynamics would only apply a phase to the wavefunction. This phase
could be read out by the phase estimation algorithm
\citep{kitaev_quantum_1995,abrams_quantum_1999,nielsen_quantum_2000},
forming the required energy black box. Although the large prefactor
of this algorithm would make it slower for small molecules than the
equivalent quantum FCI calculation, it benefits from a superior asymptotic
scaling as well as from the fact that only minimal modifications would
need to be made to insert the perturbation $\boldsymbol{\mu}$ into
the calculation. For example, simulations with different nuclear coordinates
proceed in exactly the same way, while an electromagnetic field requires
only a small modification of the simulated Hamiltonian \citep{christof_zalka_simulating_1998}.

It should be remarked that current quantum algorithms for energy estimation,
such as the ones mentioned above, rely on quantum phase estimation,
which has been criticized as inefficient
\citep{brown_limitations_2006} because its cost grows exponentially
with the number of bits of precision sought. This could be significant for
gradient estimation, which may require precise energy evaluations to avoid
numerical errors. To estimate the cost, we note that if the gradient is
desired to $n$ bits of precision (as in Eq. \ref{eq:gradient}), the black
box should evaluate the energy to \begin{equation}
n_E=\log_2 \left[\frac{E_{\mathrm{max}}-E_{\mathrm{min}}}{(mh/2^n)(\theta/2\pi)} \right]\approx n+\log_2\frac{2\pi}{\theta}\end{equation} 
bits of precision \citep{jordan_fast_2005}, where $\cos^2\theta$ is the desired success probability
of the algorithm. For example, with $\theta=\pi/8$, the algorithm succeeds
85\% of the time and requires four more digits of precision in the energy
than is desired in the gradient. The four additional digits present only a
constant overhead, meaning that the computation of any molecular property
at any precision is, up to a constant factor, as hard as computing the
energy of the same molecule at the same precision.

Finally, a limitation of current quantum simulation algorithms is
that they are generally spin-free and non-relativistic, which limits
the ability to compute derivatives such as indirect spin-spin coupling.

\section{Newton's method and geometry optimization}

Perhaps the single most common use of molecular derivatives is molecular
geometry optimization. We can therefore use it to illustrate some
of the advantages of a quantum algorithm over a classical one, including
a quantum version of Newton's method, which offers an additional quadratic
speedup over its classical counterpart.

A simple way for finding the locally optimal geometry is to perform
the standard Newton iterations, \begin{equation}
\mathbf{R}_{n+1}=\mathbf{R}_{n}-\left(\left.\frac{\text{d}E}{\text{d}\mathbf{R}}\right|_{\mathbf{R}_{n}}\right)\cdot\left(\left.\frac{\text{d}^{2}E}{\text{d}\mathbf{R}^{2}}\right|_{\mathbf{R}_{n}}\right)^{-1},\label{eq:step}\end{equation}
until convergence is reached. Here, $\mathbf{R}_{n}$ are the nuclear
coordinates at the $n^{\text{th}}$ iteration, and $\left.\frac{\text{d}E}{\text{d}\mathbf{R}}\right|_{\mathbf{R}_{n}}$
and $\left.\frac{\text{d}^{2}E}{\text{d}\mathbf{R}^{2}}\right|_{\mathbf{R}_{n}}$
are, respectively, the gradient and Hessian of $E$ with respect to
nuclear displacement (the {}``molecular gradient'' and the {}``molecular
Hessian''). If a quantum computer were used to compute the derivatives,
one would require exactly 3 calls to a black box for $E$ per iteration:
one for the gradient and two for the Hessian. A classical approach,
on the other hand, would be much slower, requiring at least $d^{2}+1$
function calls for finite difference, and approximately $O(d)$ effort
in the analytical case %
\footnote{Quasi-Newton methods (such as the Broyden\textendash{}Fletcher\textendash{}Goldfarb\textendash{}Shanno
method), or even simpler methods such as simple gradient descent,
can remove the need to compute the molecular Hessian at each step,
or at all. While such schemes are useful and widely applied, we do
not discuss them here because they are typically slower and, for a
given number of steps, less accurate than Newton's method. While they
offer a classical computational advantage, on a quantum computer that
advantage would be erased by the ability to rapidly compute the exact
Hessian at each step. %
}. For large molecules with large $d$, this savings could prove significant,
even if each energy evaluation takes much longer on a quantum machine
than on a classical computer.

There are many classical tricks available for speeding up the convergence
of Newton's method if the initial guess is not close to a local minimum,
in which case the usual Newton step might be inappropriately large.
Techniques such as trust regions and level shifts \citep{fletcher_practical_2000}
are still available to quantum computers, or they can be implemented
as classical post-processing.

In addition, we remark that Newton's method is the first in the class
of Householder methods, which offer a rate of convergence of $\ell+1$,
provided that derivatives up to order $\ell+1$ exist and can be calculated.
A quantum computer could be used to accelerate Householder methods
of any degree, requiring $\sum_{m=1}^{\ell+1}2^{m-1}=2^{\ell+1}-1$
calls to the black box for order-$\ell$ Householder optimization
method. Although exponential in $\ell$, this expression is independent
of system dimension $d$.

Of course, Newton's method is only useful for local minimization,
and we are often interested in global optimization. Here, we can use
a quantum version of the multistart technique, called the quantum
basin hopper \citep{bulger_quantum_2005-1,bulger_combining_2007,w_p_baritompa_grovers_,zhu_simulated_2009}.
A number of points is selected at random, and each is followed, using
a local search, to its local basin (if a quantum version of Newton's
method is used for the local search, such as the one we propose above,
we can get the usual quadratic convergence). Then, the minima of all
the basins are compared and the least one chosen as the global minimum.
Quantum computers could add a quadratic speed-up to such a multistart
technique, since the resulting local minima form an unstructured database
that can be searched using Grover's algorithm \citep{grover_fast_,nielsen_quantum_2000}
with a quadratic speed-up. As D{\"u}rr and H{\o}yer pointed out \citep{durr_quantum_1996},
a Grover search can find the minimum of an unstructured database with
$O(\sqrt{K}\log K)$ calls to the database (where $K$ is the number
of database entries, i.e. multistart points), as opposed to the classically
required $O(K\log K)$ queries.

\section{Conclusion}

We have shown that Jordan's quantum gradient estimation algorithm
can be applied to the estimation of time-independent, non-relativistic
molecular properties. Doing so requires a quantum electronic-structre
black box, for which known quantum simulations algorithms are well suited.
The quantum algorithm offers a speed-up from the classical
cost of $O\left(d^{\left\lfloor n/2\right\rfloor }\right)$ for analytical
derivatives to the quantum query complexity of $2^{n-1}$.
That is, the number of energy calculations required on the quantum
computer is independent of $d$, and thus of the system size, which
could offer a significant advantage for the computation of
properties of large systems. In particular, it would make the molecular
Hessian of any molecule only twice as expensive as its molecular gradient,
enabling a fast, local geometry optimization using Newton's method.
Finally, global optimization could combine the local Newton's method
with Grover search to offer an additional quadratic speed-up over the classical
multi-start technique.

\begin{acknowledgments}
We acknowledge support from the Army Research Office under contract
W911NF-07-0304. I.K. thanks the Joyce and Zlatko Balokovi{\'c} Scholarship.
\end{acknowledgments}

\end{document}